\documentclass[preprint,amsmath,amssymb,superscriptaddress]{revtex4}
\usepackage{graphicx}% Include figure files
\usepackage{dcolumn}% Align table columns on decimal point
\usepackage{subfigure}
\usepackage{rotating}

\begin{document}

%\preprint{APS/123-QED}

\title{Precision control of thermal transport in cryogenic single-crystal silicon devices}

\author{K. Rostem}
\email{karwan.rostem@nasa.gov}
\affiliation{Dept. of Physics \& Astronomy, The Johns Hopkins University, 3400 N. Charles St., Baltimore, MD 21218}
\affiliation{NASA Goddard Space Flight Center, 8800 Greenbelt Road, Greenbelt, MD 20771}
\author{D. T. Chuss}
\author{F. A. Colazo}
\author{E. J. Crowe}
\author{K. L. Denis}
\author{N. P. Lourie}
\author{S. H. Moseley}
\author{T. R. Stevenson}
\author{E. J. Wollack}
\affiliation{NASA Goddard Space Flight Center, 8800 Greenbelt Road, Greenbelt, MD 20771}

\date{\today}

\begin{abstract} 

We report on the diffusive-ballistic thermal conductance of multi-moded single-crystal silicon beams measured below 1 K. It is shown that the phonon mean-free-path $\ell$ is a strong function of the surface roughness characteristics of the beams. This effect is enhanced in diffuse beams with lengths much larger than $\ell$, even when the surface is fairly smooth, 5-10 nm rms, and the peak thermal wavelength is 0.6 $\mu$m. Resonant phonon scattering has been observed in beams with a pitted surface morphology and characteristic pit depth of 30 nm. Hence, if the surface roughness is not adequately controlled, the thermal conductance can vary significantly for diffuse beams fabricated across a wafer. In contrast, when the beam length is of order $\ell$, the conductance is dominated by ballistic transport and is effectively set by the beam area. We have demonstrated a uniformity of $\pm$8\% in fractional deviation for ballistic beams, and this deviation is largely set by the thermal conductance of diffuse beams that support the micro-electro-mechanical device and electrical leads. In addition, we have found no evidence for excess specific heat in single-crystal silicon membranes. This allows for the precise control of the device heat capacity with normal metal films. We discuss the results in the context of the design and fabrication of large-format arrays of far-infrared and millimeter wavelength cryogenic detectors. 

\end{abstract}

\maketitle

%%%%%%%%%%%%%%%%%%%%%%%%%%%%%%
%%%%%%%%%%%%%%%%%%%%%%%%%%%%%%
\section{Introduction}

The management of heat in dielectric beams and membranes is an important part of the design in micro-electro-mechanical systems (MEMS)~\cite{Savin,Maldovan,Yefremenko1}. While the physics of thermal transport in dielectric materials is well understood, the effect of fabrication processes on the surface physics is less clear and widely treated as a hidden variable in the evaluation of the device performance. This uncertainty can prolong the design-test cycle, where a parameter such as beam geometry is iterated until the target thermal conductance $G$ is obtained~\cite{ACTPol,SPT90GHz}. 

\begin{figure}[htbp]
\begin{center}
\includegraphics[width=8.5cm]{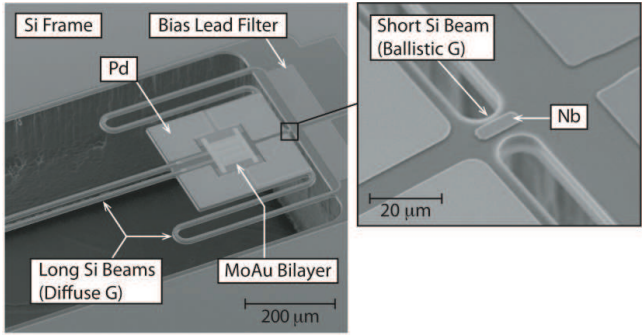}
\caption{SEM image of a TES detector fabricated for a 40 GHz focal plane~\cite{Rostem}. The single-crystal silicon membrane volume is $5\times350\times450$ $\mu$m$^3$. The thermal conductance from the membrane to the frame is effectively set by the ballistic-dominated thermal conductance of the short beam (magnified). The long silicon beams are $5\times15\times785$ $\mu$m$^3$, and merely support TES bias and rf leads. The thermal conductance of the long silicon beams is diffusive-ballistic and sub-dominant to the conductance of the ballistic beam. The Pd layer deposited on the membrane sets the detector heat capacity. }%%%%%%%
\label{fig:Q-band-detector}
\end{center}
\end{figure}

In our application, the MEMS is a cryogenic detector known as the Transition-Edge Sensor (TES)~\cite{TESIrwinHilton}. Figure~\ref{fig:Q-band-detector} shows a TES designed for measurements of the polarization state of the Cosmic Microwave Background~\cite{Rostem,Eimer}. The TES is comprised of a superconducting MoAu bilayer deposited on a single-crystal silicon substrate. The silicon is etched around the bilayer and other electrical components to form a membrane that is supported by dielectric beams. The total thermal conductance $G(T)$ of the beams is a critical parameter that determines the sensitivity ($\propto \sqrt{G}$), response time ($\propto 1/G$), and saturation power of the TES detector ($\propto \int G\,dT$) (saturation energy in the case of a TES micro-calorimeter)~\cite{TESIrwinHilton}. The response time is also a function of the total heat capacity $C(T)$ of the detector, which is the sum of the heat capacity of the silicon membrane and normal metal films.

At temperatures far below the Debye temperature of the lattice ($\Theta_{Si}=645$ K), anharmonic phonon collisions in the bulk of the dielectric are negligible, and the phonon-mean-free path $\ell$ is restricted by the cross sectional dimensions of the beam~\cite{Kittel}. In state-of-the-art far-infrared and millimeter TES detectors operating at sub-Kelvin temperatures~\cite{Khosropanah,ACTPol,SPT90GHz}, the beam length $L$ is much larger than $\ell$, and the thermal conductance is often described by the Fourier law for phonon diffusion, $G(T) = \kappa(T)\,wt/L$, where $w$ and $t$ are the width and thickness of the beam, and $\kappa(T)$ is the bulk thermal conductivity of the dielectric. $\kappa(T)$ is determined empirically with the fabrication and testing of numerous beams of varying width and length, and the aspect ratio $w/L$ is tuned to achieve the desired conductance. However, this approach rarely provides a complete view of the thermal physics, and higher order effects such as surface roughness of the beams have been identified as the likely cause of the reported non-uniformity in $G$ across detector arrays~\cite{ACTPol,SPT90GHz}. 

A simple analysis can readily show the sensitivity of phonon propagation in long beams ($L\gg\ell$) to the details of the surface physics. For a random Gaussian surface, the roughness can be characterized by the ratio of the standard deviation $\sigma$ of the surface features to the peak thermal wavelength $\lambda_{th}(T)$ of the incident phonon field. When $\sigma/\lambda_{th} \ll 1$, phonons are reflected specularly from the surface~\cite{Beckmann}. As the wavelength becomes much smaller than the scale size of the surface, $\sigma/\lambda_{th}\gg1$, phonons are scattered over all angles, or diffused. Assuming isotropic diffusion, it can be shown that for a rectangular beam, the phonon mean-free-path $\ell$ scales logarithmically with $r = w/t$ ($\ell\approx0.75\,wr^{-1}\ln2r$ for $r>1$~\cite{Wybourne}). Hence, in a beam where $L\gg\ell$, phonons experience a large number of scattering events, $N\sim L/\ell$, and the heat flux entering the beam at one end $P_0(T)$ is reduced to $P(T) = P_0(T)\,(1-f)^N$ at the exit port, where $f$ is the diffuse fraction. The derivative $dG/dN$ is a measure of the sensitivity of $G$ to changes in $N$ and proportional to the power-law function $(1-f)^N$. 

In practice, the statistics of the surface roughness can be heavily biased by the fabrication and is likely non-Gaussian. Dielectric beams can be roughened in various forms depending on the compatibility and properties of the etchants with the dielectric material. Preliminary tests can shed light on the roughening steps, however, fabrication processes can be difficult to control in practice. Details such as base pressure and process chamber conditioning during dry-etch steps can vary between batches of devices, especially in shared facilities. As a result, the characteristics of the roughness of a single surface can change significantly over spatial scales shorter than the beam length, and between wafers. Furthermore, not all surfaces of a beam are rough to the same degree. These effects add to the unpredictable variability in $N$ and therefore $G$ in beams dominated by diffusive thermal transport. 

The work in this paper is focused on understanding the effects of fabrication on the thermal conductance in single-crystal silicon beams. The beam geometries explored span several orders of magnitude in $L/wt$. The results demonstrate that $G$ in short ($L\lesssim10$ $\mu$m, $L/\ell\le$ 1) uniform beams with ballistic-dominated phonon transport is insensitive to the fabrication conditions and can be realized with precision in devices on different wafers. This approach is suitable for TES detector arrays in ground-based and air-borne telescopes, where the detector noise equivalent power requirements is greater than 10$^{-18}$ W/$\sqrt{\rm Hz}$. 

Throughout this paper, the terms \emph{diffuse} and \emph{ballistic} are used to describe the scale size of the phonon transport in a silicon beam. In the presence of surface roughness, the thermal transport in a long beam with $L/wt\gg1$ is dominated by diffuse reflections off the beam walls, and $L/\ell\gg1$. For a short beam, ballistic transport dominates and $L/\ell\le1$. We emphasize that these limits bound the diffusive-ballistic conductance observed in practice.  

In Sec.~\ref{sec:method}, we briefly describe the device fabrication and readout hardware employed for the low-temperature measurements of $G$ and $C$. In Sec.~\ref{sec:model}, we describe the theory of radiative phonon transport in diffusive-ballistic beams. In Sec.~\ref{sec:diff-ball}, we present results that illustrate the effect of fabrication on the diffusive-ballistic nature of the thermal conductance. In Sec.~\ref{sec:controlG}, we describe the superior performance of ballistic beams for the control of $G$ of TES detectors. Finally, we demonstrate that because of the low specific heat capacity of single-crystal silicon, the heat capacity of the detectors fabricated on this substrate can be effectively controlled using a normal metal film with high specific heat. 

%%%%%%%%%%%%%%%%%%%%%%%%%%%%%%
%%%%%%%%%%%%%%%%%%%%%%%%%%%%%%
\section{Methodology\label{sec:method}}

\subsection{Fabrication}

The fabrication of the TES detectors reported in this paper is described in detail in Ref.~\cite{Denis}, but for completeness, we provide a brief description here. Fabrication starts with a float-zone-refined p-type (100) boron doped silicon-on-insulator wafer. Heat flow is primarily in the [110] direction. The single-crystal silicon layer is 5 $\mu$m thick, with a room temperature resistivity of 5 k$\Omega$-cm and a carrier density of 5$\times$10$^{11}$ cm$^{-3}$. A MoAu bilayer is deposited by a combination of dc plasma sputter deposition and electron beam evaporation. The Mo and Au are photolithographically patterned to define the superconducting sensor. The Au layer is etched by ion milling and the Mo layer is etched by reactive ion etching (RIE). Further metallic components are deposited on the device layer to enable the required operation of the pixel at rf frequencies, and to enable on-chip diagnostic capabilities such Johnson noise thermometry as needed. Au resistors are deposited by electron beam evaporation and patterned by lift-off. Following an in-situ rf Ar plasma cleaning of the wafer, a Nb microstrip layer is sputter deposited, photolithographically patterned, and etched in fluorine-based RIE chemistry. The silicon membrane is defined and reactive ion etched down to the buried oxide layer. The silicon wafer is then bonded to a pyrex handle wafer using a sacrificial wax. The backside of the silicon wafer is patterned by photolithography and the bulk silicon is etched by inductively-coupled plasma deep reactive-ion etching (DRIE). The wax is then dissolved in solvent to release the devices. 

\subsection{Experimental Hardware}

Measurements were performed in a mechanically-cooled system that reaches 3 K at the coldest stage. Further cooling of the detector stage down to 100 mK was achieved using an adiabatic demagnetization refrigerator. The devices were mounted and thermally anchored to a gold-plated copper package, and wirebonded to a Superconducting Quantum Interference Device (SQUID) cold stage amplifier for Johnson noise current measurements. Further amplification was achieved with a SQUID series array mounted at the 3 K stage. The temperature of the 100 mK cold stage was read with a calibrated RuO$_x$ thermometer~\cite{ruox-therm} with calibration traceable to the National Institute of Standards and Technology. The precision of the temperature control was 32 $\mu$K rms at 150 mK.  

The room temperature electronics was comprised of a custom low-noise pre-amplifier, feedback electronics for flux-locked-loop operation, commercial low-noise high-gain amplifier and low-pass filter (SRS-650), and commercial 16 Bit digital-to-analog electronics (NI PCI-6120) for digitization of the amplified feedback signal. 

\subsection{Johnson Noise Thermometry}

The method of Johnson noise (JN) thermometry is frequently utilized for the characterization of the thermal properties of thin-films below 1 K~\cite{DC-SQUID}. At audio frequencies (0.5-20 kHz) and sub-Kelvin temperatures, the current noise spectral density of a resistor $R$ in series with the SQUID input coil takes the form $S_I (f) = {4k_BT_e}\,[1+(f/f_c)^2]^{-1}/R$, where $T_e$ is the electron temperature, $f_c = R/(2\pi L)$ is the characteristic frequency of the circuit, and $L$ is the total inductance of the input coil and stray inductance in the input circuit loop. This equation directly relates measurements of $S_I$ to $T_e$. For the test devices that are described in Ref.~\cite{Denis}, the JN thermometer and heater are 240$\times$230$\times$0.27 $\mu$m$^3$ Au resistors deposited on top of interdigitated Nb leads. The Kapitza conductance at the Au-silicon interface and the electron-phonon conductance in the Au are approximately 10 nW/K each at 100 mK~\cite{Swartz, RevModPhys.78.217}. These conductances are a factor of 50 greater than the highest silicon beam conductance measured in our devices. We therefore assume the electron system in the thermometer is in thermal equilibrium with the phonons in the silicon membrane. The maximum expected temperature gradient across the membranes is less 0.5 mK across the 0.1-1 K temperature range. 

The random error in a temperature measurement from the JN current can be estimated from the radiometer equation $\sigma_T = T/\sqrt{B\,\tau_m}$~\cite{Dicke,DC-SQUID}, where $B=5$ kHz is the bandwidth over which the current noise power spectral density was measured and $\tau_m = 102$ s is the measurement time. Hence, the fractional error throughout the temperature range of our measurements is less than 0.5\%. Measurements of the JN current as a function of bath temperature were used to fully calibrate the cryogenic and warm readout amplifiers. Typical readout noise of 15 pA/$\sqrt{\rm Hz}$ referenced to the input circuit was achieved. The JN signal was 30-100 pA/$\sqrt{\rm Hz}$ in the range 0.1-1 K. 

To measure $G$ of the silicon beams, a small signal dc heater power was applied to the membrane and the increase in temperature was recorded. The relationship between heater power and membrane temperature at a given bath temperature $T_B$ can be parameterized by the power law function $P=K(T^n-T_B^n)$, where $P$ is the heater power, $T$ is the membrane temperature, $K$ is a constant that is proportional to the number of phonon modes, and $n$ is the temperature exponent that is indicative of the temperature dependence of the phonon mean-free-path. The first order derivative yields the thermal conductance $G(T)=nKT^{n-1}$. In our measurements, the change in temperature was small, $\sim$10\%, and the approximation $G(T) \cong \Delta P(T)/\Delta T$ was valid. 

\subsection{TES Current-Voltage Thermometry}	

We have also utilized TES current-voltage curves to determine the thermal conductance of detectors in which normal metal films for JN thermometry were omitted. When a TES is voltage-biased on the normal-to-superconducting phase transition, the silicon membrane temperature is fixed at the critical temperature of the MoAu bilayer, $T_C\cong150$ mK. The thermal conductance is then determined from a fit to the Joule power dissipated in the TES as a function of bath temperature. In this case, $P=P_J=K(T_C^n-T_B^n)$ and $G(T_B)=nKT_B^{n-1}$, where the parameters $K$ and $n$ are equivalent to those derived from JN thermometry in the same temperature range. 

%%%%%%%%%%%%%%%%%%%%%%%%%%%%%%
%%%%%%%%%%%%%%%%%%%%%%%%%%%%%%
\section{Modeling Heat Transport\label{sec:model}}

In building a model of heat flow in a silicon beam, it is important to first identify the thermal baths that define the boundary conditions in the model. A thermal bath is ideally a blackbody that provides a matched termination for phonons at all wavelengths. In a detector such as that shown in Fig.~\ref{fig:Q-band-detector}, each silicon beam is connected to two thermal baths, the frame and the membrane. The electron system in the Pd film is strongly coupled to the phonon system in the membrane. At 150 mK, the electron-phonon conductance in the Pd is 30 nW/K~\cite{hot-e-Pd}, and the Kaptiza conductance between the Pd and the silicon membrane is $\sim$ 130 nW/K~\cite{Swartz}. For comparison, the highest beam conductance measured is 240 pW/K at 150 mK. The silicon frame is effectively a blackbody since its lateral dimensions are much larger than the peak thermal wavelength. At 100 mK, the peak thermal wavelength is $\lambda_{th}\sim 4hc/k_BT=0.6$ $\mu$m in silicon, where $h$ and $k_B$ have the usual definitions and $c=5070$ m s$^{-1}$ is the speed of sound for transverse waves. A phonon scattered from a beam into the silicon frame has a low probability of coherently backscattering into the beam. Given the 5 $\mu$m thickness of the silicon layer, and the thermal wavelengths above 100 mK ($\lambda_{th}\approx0.6\,(0.1/T)$ $\mu$m), the membrane and the frame can be modeled as three-dimensional multi-moded thermal baths, where the number of modes is given by the Planck distribution and is much larger than unity. 

The beams behave as three-dimensional multi-moded acoustic waveguides since the widths ($w>13$ $\mu$m) and 5 $\mu$m thickness are much larger than $\lambda_{th}$ at 0.1-1 K. Phonons emitted by the baths couple radiatively to the silicon beams, propagate unimpeded in the bulk of the beams, and scatter at the beam surfaces. When the surfaces of the beams are perfectly smooth, phonons coupled to the beams are scattered elastically without any loss of coherence in the beams. The conductance in this case is determined by the beam areas and is referred to as ballistic. In the case of rough surfaces, phonons scatter diffusively with a large probability of backscatter. The scattering is elastic but the phonon wavevectors are randomized at the surface. If the surface features form cavities much larger than $\lambda_{th}$, the surface can be thought of as a blackbody.  The conductance is thus boundary-limited and was first described theoretically by Casimir~\cite{Casimir}. An important assumption in Casimir's formalism is the isotropic (Lambertian) reemission of phonons by the beam surfaces.

\begin{figure}[htbp]
\begin{center}
\includegraphics[width=8.5cm]{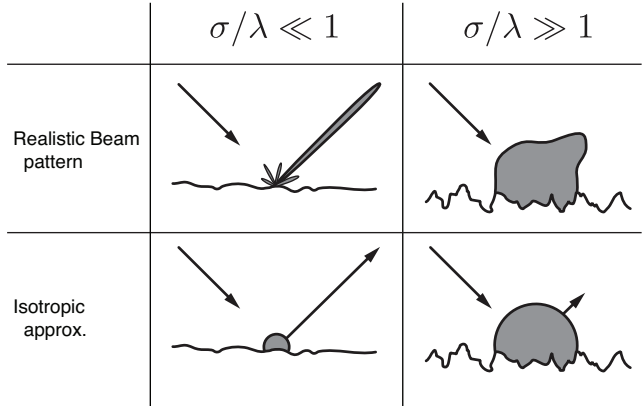}
\caption{Beam patterns of phonons reflecting off a fairly smooth ($\sigma/\lambda\ll1$) and a very rough surface ($\sigma/\lambda\gg1$). The diffuse fraction of the realistic beam patterns are generally approximated as isotropic (Lambertian), which is relatively simple to integrate into models of thermal transport.  }
\label{fig:beamPattern}
\end{center}
\end{figure}

In practice, the scattered pattern of phonons from a rough surface is strongly dependent on the surface characteristics. When the roughness can be described statistically, the reflection pattern is averaged over the ensemble of the surface realizations. For a surface with a Gaussian correlation function, the roughness can be characterized by the correlation length $\xi$ and standard deviation $\sigma$ of the surface features above a mean height. Figure~\ref{fig:beamPattern} illustrates the scattered pattern as a function of $\sigma/\lambda$~\cite{Beckmann}. For a nearly smooth surface, $\sigma/\lambda\ll1$, phonons are largely reflected in the specular direction. When $\sigma/\lambda\gg1$, phonons incident on the surface are diffused in all directions, however, the flux is not Lambertian. Nonetheless, in numerous theoretical studies of boundary-limited scattering~\cite{Casimir,Berman,Curdy,Wybourne,Klitsner}, the reflected pattern is decomposed into a diffuse Lambertian fraction $f$ and a specular fraction $1-f$, as illustrated in Fig.~\ref{fig:beamPattern}. The diffuse reflection is modeled as inelastic scattering that redistributes the phonon wavevector, although strictly speaking all reflections are elastic at the surface facets that define the roughness. The surface can then be thought of as a blackbody with emittance $f$. Hence, a completely lossless but rough surface can be modeled as a lossy surface~\cite{Berman,Curdy,Wybourne,Klitsner}. This approach is numerically less intensive than calculations of the reflection pattern based on the exact surface profile, or the statistics of the surface~\cite{Beckmann}. % (e.g. Eq.~\ref{eqn:GaussSurf})

%One could add this section to the text
%The power reflection coefficient is given by~\cite{Beckmann}
%\begin{eqnarray}
%\langle R\rangle = \exp\left(-a\,\frac{\sigma^2}{\lambda^2}\right)\,\left[1+\pi\frac{\xi^2}{A}b\sum_{m=1}^\infty \left(-a\,\frac{\sigma^2}{\lambda^2}\right)^m\frac{1}{m!\,m}\exp\left(-\frac{c}{m}\,\frac{\xi^2}{\lambda^2}\right)\right], 
%\label{eqn:GaussSurf}
%\end{eqnarray}
%where $a=4\pi^2 (\cos\theta_i+\cos\theta_r)^2$, $A$ is the total area of the reflecting surface, $\theta_i$ is the angle of incidence, and $\theta_r$ is the angle of reflectance. The geometric functions $b$ and $c$ are dependent on $\theta_i$, $\theta_r$, and the angle between the planes of incidence and reflection. In deriving Eq.~\ref{eqn:GaussSurf}, it is assumed $\xi/\lambda$ and $A/\xi^2$ are much greater than unity.

\begin{figure}[!t]
\begin{center}
\includegraphics[width=8.5cm]{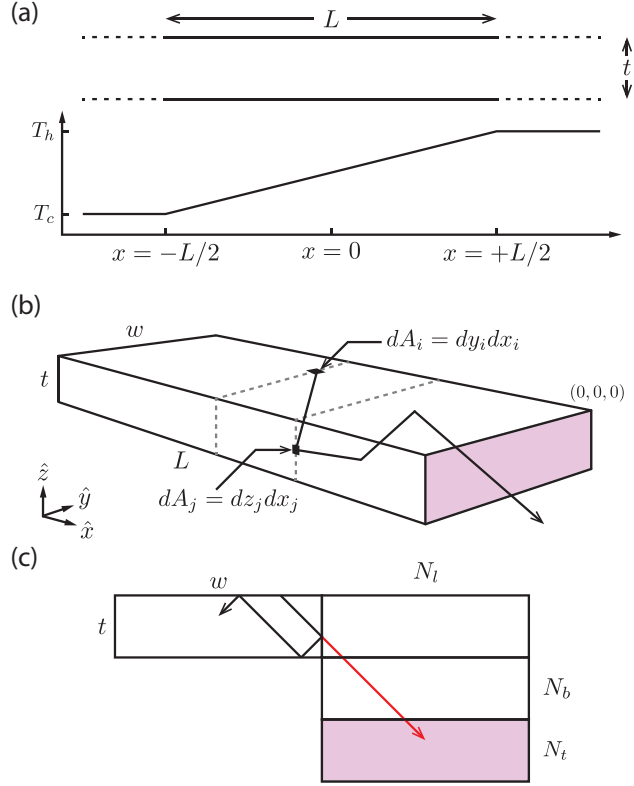}
\caption{(Color online) (a) Beam geometry and linear temperature profile assumed in the model of heat transport. The temperature profile is linear across the beam. The temperature is constant at the thermal baths, which are represented by diffuse surfaces ($f=1$) that extend to distances much larger than $L$. (b) Each surface element of the beam radiates phonons into all solid angles. The radiated flux from the infinitesimal area $dA_i$ that reaches the shaded beam port is reduced depending on the diffuse fraction of the beam surfaces. (c) Using the method of images, the path of specularly scattered phonons can be re-imaged to one vector. The components of the vector map the number of scattering events at each surface of the beam. }
\label{fig:ModelDiagram}
\end{center}
\end{figure}

Our model of heat flow in a silicon beam has the following assumptions: (i) The ends of the beam constitute thermal baths with constant temperature and unity diffuse fraction. The baths continue to distances much larger than the beam length as shown in Fig.~\ref{fig:ModelDiagram}(a). This construct is numerically convenient from the perspective of the heat transport model presented, and is similar to that adopted in Ref.~\cite{Curdy}. (ii) The surfaces of the beam reflect phonons with a diffuse fraction $f$ that is isotropic, and a specular fraction given by $1-f$. (iii) The gradient of the temperature profile across a beam is linearly dependent on $f$ and constant as a function of length. This assumption is valid in multi-moded beams, where the number of modes is much larger than unity and the diffuse phonon reflection can be treated as diffuse emission~\cite{Casimir}. Models that iteratively solve for the temperature profile in multi-moded beams show negligible deviations from a constant temperature gradient~\cite{Klitsner}. When the phonon mean-free-path is much smaller than the beam length, or $f$ is close to unity, phonon diffusion is the dominant term of the total heat flux, and the constant temperature gradient is an excellent approximation. The approximation also holds when $f$ approaches zero, or the mean-free-path is comparable to the beam length, since the temperature gradient is linearly dependent on $f$ (the mathematical description of this detail is in the Appendix). (iv) The phonon modes that propagate are two orthogonal transverse modes with a sound speed of $v_t=5070$ m/s, and one dilatational mode with a sound speed of $v_l=13600$ m/s. Each mode has a thermal spectrum. (v) Phonons propagate unimpeded in the crystalline bulk of the silicon beams, and only scatter at the beam surfaces. (vi) The diffuse flux is reemitted according to Lambert's law of thermal radiation. (vii) All modes scatter diffusively with the same probability $f$, and there is no spectral dependence for $f$. 

It is possible to map $f(\lambda_{th})$ by fitting measurements of $G(T)$ and calculating the peak thermal wavelength as a function of $T$~\cite{Klitsner}. For a Gaussian surface~\cite{Beckmann}, when $\sigma/\lambda\ll1$, the specular power reflection coefficient is proportional to $\lambda^{-2}$ and the diffuse coefficient to $\lambda^{-4}$; when $\sigma/\lambda\gg1$ the reflection coefficient is independent of wavelength. Hence, the thermal spectrum can be scattered very differently as temperature changes, and in recognizing the limitations of our model, we do not extrapolate the results to $f(\lambda_{th})$.

Figure~\ref{fig:ModelDiagram}(b) illustrates the radiative model. The beam is divided into infinitesimal areas $dA_i=dy_i\,dx_i$ that radiate heat into all solid angles. The phonons are scattered several times before reaching the beam exit port and thermalizing in a bath. During each scattering event, the phonon flux is reduced by a factor $f_k$ that is the diffuse fraction of the surface $k$. The nature of the problem is then to simultaneously calculate the flux that is viewed by $dA_j$ from $dA_i$ and count the number of scattering events from each surface of the beam. For the latter, the complex path of the phonons can be unfolded, or re-imaged, to a single vector that reaches an equivalent beam exit port as shown in Fig.~\ref{fig:ModelDiagram}(c). The components of the vector then equate to the number of scattering events from each surface. 

The details of the radiative model are described in the Appendix~\ref{sec:app:A}. We define an equivalent view factor for a rectangular waveguide 
\begin{eqnarray}
F(x_i) = \sum_{m=1}^4 \left[ F^\parallel (f_m; x_i) + \sum_{s=1}^2 F^\perp (s, f_m; x_i) + F^D(f_m; x_i) \right],
\label{eqn:Fx-final}
\end{eqnarray}
where $F^\parallel (f_m; x_i)$ and $F^\perp (s, f_m; x_i)$ are equivalent view factors between parallel and perpendicular surfaces of the beam respectively, and $F^D(f_m; x_i)$ is the contribution of the direct (unscattered) flux from the surfaces of the beam to the cross sectional area at $x=0$ (reference plane). $f_m$ is the diffuse fraction of the surface emitting the flux, and the index $m$ runs over all surfaces of the beam. The index $s$ runs over the surfaces perpendicular to $m$. Included in the terms $F^\parallel (f_m; x_i)$ and $F^\perp (s, f_m; x_i)$ is a factor $\prod_{k=1}^4 (1-f_k)^{N_k}$, where $N_k$ is the total number of scattering events from each surface as described in Fig.~\ref{fig:ModelDiagram}(c). For the baths, $F_B(x_i)$ is obtained by setting $f_m=1$ in Eq.~\ref{eqn:Fx-final}. The total heat flux through the beam is $P\equiv P|_{x=0}$,
\begin{eqnarray}
P &=& q \Bigg[ \int_{-L/2}^0 T^4(x_i)\,F(x_i)\,-\,\int_0^{L/2} T^4(x_i)\,F(x_i)  + \nonumber \\
	    &\,\,& T_c^4\int_{-\infty}^{-L} F_B(x_i)\,-\,T_h^4\int_L^\infty F_B(x_i) \Bigg]dx_i, 
\label{eqn:Px0}
\end{eqnarray}
where $q = \pi^2k_B^4/(120\hbar^3\bar{v}^2)$, and $\bar{v}=\sqrt{2/v_t^2+1/v_l^2}$ is an average group velocity. Given the linear temperature profile $T(x_i)=x_i\,(dT/dx_i) + T_c$, the approximation $T_h^4-T_c^4\approx4(dT/dx_i)LT^3$ where $T$ is the average of the bath temperatures, and the relation $F(x_i)\equiv F(-x_i)$, Eq.~\ref{eqn:Px0} reduces to
\begin{eqnarray}
P &=& 4q\,T^3\frac{dT}{dx_i}\left(2\int_{0}^{L/2} x_iF(x_i)\,dx_i + L\int_{L}^{\infty} F_B(x_i)\,dx_i\right),  \nonumber \\
	    &=& 4q\,T^3\frac{dT}{dx_i}\left(2I_d+LI_s\right),	
\label{eqn:Px01}
\end{eqnarray}
where $I_d$ is the net diffuse power from the beam surfaces, and $I_s$ is the net specular power that is exchanged between the baths. The conductance of the beam as a function of temperature when scaled by the maximum (ballistic) conductance $G_B=4qT^3A$ is
\begin{eqnarray}
g = \frac{G}{G_B} &=& \left(\frac{P}{A\,(dT/dx_i)}\frac{A}{L}\right) \frac{1}{G_B}, \nonumber \\
		      &=& \frac{1}{A}\left(\frac{2I_d}{L} + {I_s}\right). 
\end{eqnarray}
In the short beam limit, $I_d\rightarrow 0$, and $I_s\rightarrow A$. For non-zero diffuse fractions and as the beam length becomes much larger than the cross-sectional dimensions, $I_s\rightarrow 0$, and using $\kappa = C_V\bar{v}\ell/3$ where $C_V=16qT^3/\bar{v}$ is the volume specific heat of the material, the mean free path takes the form $\ell = 3I_d/(2A)$. 

When the diffuse fractions of the surfaces of the rectangular beam are different, all terms in the summation over the index $m$ in Eq.~\ref{eqn:Fx-final} must be evaluated. The numerical evaluation of the problem is reduced by a factor of two if there is symmetry in the diffuse fraction of the surfaces, $f_{\rm Top}=f_{\rm Bottom}$ and $f_{\rm Left}=f_{\rm Right}$. For the silicon beams discussed in this paper, the top surface and sidewalls are diffuse, and the bottom surface is specular. It is important to note that the diffuse fraction of the top and sidewall surfaces are uncorrelated. 

\begin{figure}[!t]
\begin{center}
\includegraphics[width=8.5cm]{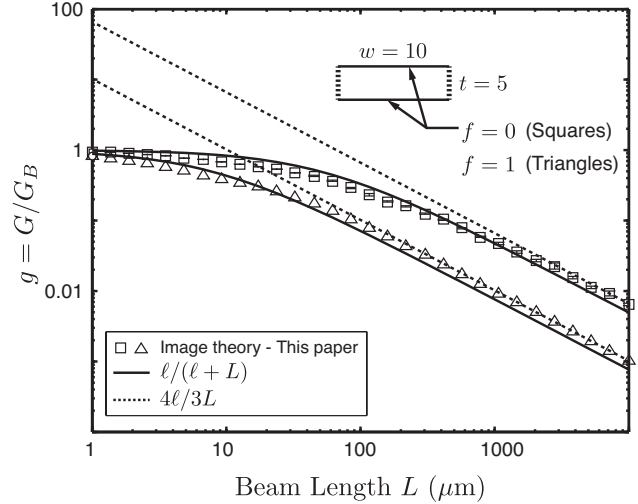}
\caption{Simulations of a beam with fixed cross-section compared to the analytic expression in Ref.~\cite{Wybourne}. The sidewall diffuse fraction is unity. The standard error from the Monte-Carlo integration is shown for each simulation. }
\label{fig:modelSimulation}
\end{center}
\end{figure}

The terms $F^\parallel (f_m; x_i)$ and $F^\perp (s, f_m; x_i)$ in Eq.~\ref{eqn:Fx-final} are evaluated using Monte-Carlo integration with 10$^5$ uniformly distributed random points averaged 10$^2$ times. $F^D(f_m; x_i)$ is evaluated with an adaptive quadrature routine. Figure~\ref{fig:modelSimulation} illustrates the accuracy of the model compared to an analytic expression for the mean-free-path $\ell$ in a rectangular beam~\cite{Wybourne}. At the diffuse limit ($L/\ell\gg1$), the conductance scales as $1/L$, and at the ballistic limit, the conductance is independent of length. The expression $g = \ell/(L+\ell)$ is remarkably accurate over the range of beam length explored in Fig.~\ref{fig:modelSimulation}. This fact highlights the importance of the phonon mean-free-path, that a single parameter can capture a significant fraction of the heat transport in the ballistic-diffusive limit. 

%%%%%%%%%%%%%%%%%%%%%%%%%%%%%%
%%%%%%%%%%%%%%%%%%%%%%%%%%%%%%
\section{Diffusive-Ballistic Conductance\label{sec:diff-ball}}	

\begin{figure}[htbp]
\begin{center}
\includegraphics[width=8.5cm]{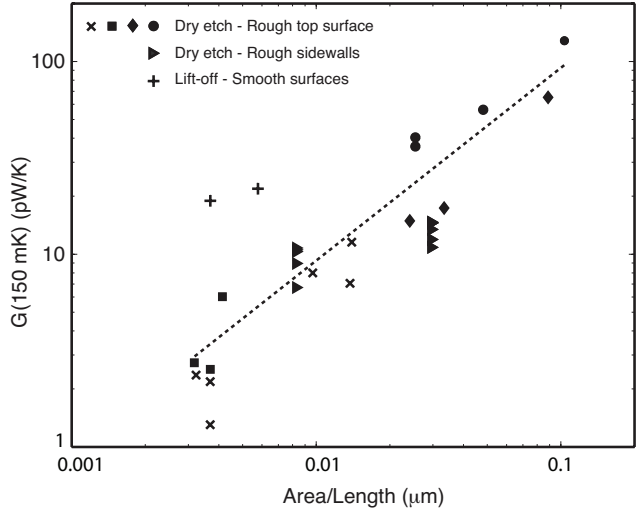}
\caption{Conductance of single-crystal silicon beams as a function of beam area to length ratio. The conductance can vary significantly depending on the fabrication conditions. Devices that were dry etched during fabrication had roughened silicon surfaces. The dashed line is the fit $G=\kappa\,A/L$ (excluding the lift-off points), and is the simplest interpretation of the data that is intended to merely guide the eye. Details of this data are summarized in Table~\ref{tbl:Data}. }
\label{fig:G-AoL}
\end{center}
\end{figure}

Figure~\ref{fig:G-AoL} shows the thermal conductance of silicon beams as a function of beam area-to-length ratio. The cause of the variability in conductance (a factor of $\sim$5 for the same beam geometry) is the roughening of the silicon beam surfaces during fabrication. The top surface of the beams are primarily roughened during a dry etch step that patterns a Nb layer. The plasma chemistry, 75\% CF4, 15\% O$_2$, etches silicon at a higher etch rate than Nb. The exact processes that roughen the silicon are unclear, but the likely candidates are micro-masking due to non-volatile species in the plasma, and slower etch of native and in-situ NbO$_x$ in the fluorine chemistry. The use of an etch stop such as 100 nm thick AlO$_x$ was successfully employed in early devices, however, these detectors exhibited very large uncontrolled heat capacity due to the amorphous nature of the etch stop~\cite{AlOx-Kinnunen}. Test devices were also fabricated with a lift-off process that left the beam surfaces smooth. The conductance of these beams were higher, suggesting boundary-limited phonon propagation as the cause of the variability in conductance of the beams with rough surfaces.

We have explored the relationship between thermal conductance and surface roughness with an extensive dataset of atomic force microscopy (AFM) images. Non-destructive AFM of a silicon beam surface a millimeter in length is not practical. Instead, the silicon device layer was imaged on the handle wafer, in close proximity to each device. For a silicon surface with a Gaussian roughness, $\sigma\sim5-20$ nm and $\xi\sim100-200$ nm depending on the processing conditions. However, we have found that a large fraction of the roughened surfaces did not satisfy the mathematical assumptions required to apply the theory of phonon scattering by nearly-smooth Gaussian surfaces, and the measured thermal conductance differed from expectations of such a theory, even when $\sigma$ was much less than the typical phonon wavelength at the temperatures explored (0.1-1 K).  We observed power spectral densities of surface height that were approximately power laws over the range of observable wavevector values. In this case, we found an empirical relationship between the power-spectral-density $S(k)$ of the surface profile and the measured magnitude and temperature dependence of the phonon mean-free-path in our silicon beams, $\ell = \pi t/(4k^2\sqrt{S(k)/(2\pi)})$, where $t$ is the silicon beam thickness and $k=0.010\,T/({\rm nm\,K})$ is the phonon wavevector. As expected from Beckmann's theoretical treatment of scattering from non-Gaussian surfaces~\cite{Beckmann-nonGauss}, the observed relationship involves the correlation function of the local tilt angle of the surface facets, $\theta_{rms}(k)=k^2\sqrt{S(k)/(2\pi)}$.   

It is challenging to precisely predict the phonon mean-free-path from information contained in the AFM data due to practical sampling considerations. This occurs for several reasons: The statistics of the surface roughness are observed to vary significantly across the wafer and are sparsely sampled in $8\times8$ $\mu$m$^2$ patches set by the resolution and stability of the AFM. In addition, for a multi-moded beam with thickness of order the width, details such as surface roughness of the beam sidewalls are significant, as is shown in Sec.~\ref{sec:controlG}. Non-destructive AFM imaging of the sidewall geometry is extremely difficult to perform, and such a task becomes impractical when considering the number of silicon beams that constitutes a detector imaging array. However, we have found that AFM images can provide qualitative insight to the thermal conductance in beams where the surface roughness relative to the thermal wavelength has a significant amplitude and is non-Gaussian.

\begin{figure}[htbp]
\begin{center}
\includegraphics[width=8.5cm]{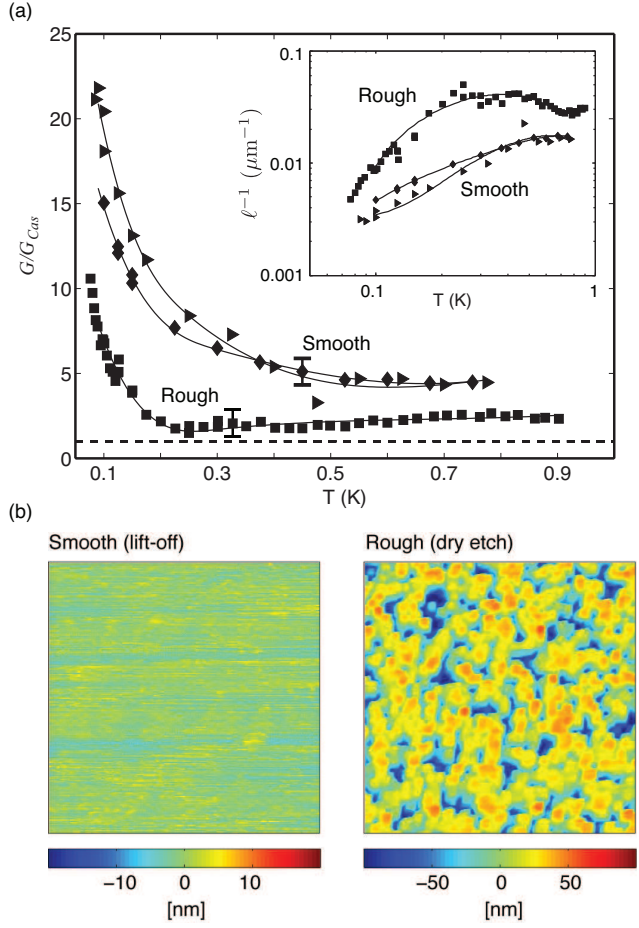}
\caption{(Color online) (a) Conductance of silicon beams scaled by the boundary-limited conductance in a rectangular beam (dashed line)~\cite{Wybourne}. Solid lines are spline fits to guide the eye. The labels refer to the top surface of the beams, which is roughened by a dry etch step during fabrication (squares), but is left smooth when the dry etch step is replaced with a lift-off process (diamonds, triangles). The inset shows the inverse phonon mean-free-path as a function of temperature. The peak in $\ell^{-1}$ suggests resonant phonon scattering from the rough surface. (b) AFM images of 6.4$\times$6.4 $\mu$m$^2$ patches representative of the top surface of the beams described in (a). The scale size estimated from the peaks in $\ell^{-1}$ is consistent with the roughness scales measured from these images. The rough surface is heavily pitted, with an average pit depth of 30 nm. The large distribution in pit size and shape is the likely cause of the observed broad resonance. Since the scale size is much smaller for the smooth surfaces ($\sigma\simeq3$ nm), the onset of the peaks occur at a much higher temperature (shorter thermal wavelengths). The striping in the AFM image of the smooth surface is due to residue and accumulation of particulates on the AFM tip. Beam dimension are $L=\{0.4,1.5,3\}$ mm, and $w=\{24,20,25\}$ $\mu$m for devices labelled with squares, diamonds, and triangles respectively.}
\label{fig:Cas-Lim}
\end{center}
\end{figure}

Figure~\ref{fig:Cas-Lim}(a) shows the conductance of smooth (lift-off) and roughened (dry etch) silicon beams when compared to the boundary-limited conductance in each case~\cite{Wybourne}. The phonon propagation transitions from diffuse to ballistic at 300 mK, where the peak thermal wavelength is $\lambda_{th}\sim200$ nm.  The minimum in conductance suggests the transition is resonant for the rough beam~\cite{Klitsner-residue}. The inset in Fig.~\ref{fig:Cas-Lim}(a) shows the inverse phonon mean free path $\ell^{-1}$. The scale height of the scatterers $\zeta$ can be estimated from the temperature at which $\ell^{-1}$ peaks. Using the Rayleigh criterion, $\zeta\approx\lambda_{th}/8\sin \gamma$~\cite{Beckmann}, where $\lambda_{th}$ is the phonon wavelength at the peak temperature, and $\gamma$ is the phonon angle of incidence to the surface. Assuming an average over all angles, $\langle \sin\gamma \rangle=2/\pi$, the Rayleigh criterion suggests $\zeta\simeq40$ nm for the rough beam and $\zeta\lesssim6$ nm for the smooth beams. These estimates are consistent with the roughness measured from the AFM images shown in Figure~\ref{fig:Cas-Lim}(b). The rough (dry etch) surface is heavily pitted, with an average pit depth of 30 nm and average size of 250 nm that is comparable to $\lambda_{th}$ at 300 mK. The scale size is much smaller for the smooth (lift-off) surfaces, $\sigma\simeq3$ nm, which leads to the onset of resonance at a much higher temperature. 

In addition to phonon scattering by rough surfaces of a beam, organic residue such as photoresist on the surfaces can enhance diffuse scattering. Thin layers of metals and non-metals hundreds of Angstroms thick have been shown to induce diffuse transport in silicon~\cite{Klitsner-residue}. In our devices, the reduction and variability in $G$ in a few cases were at least qualitatively attributed to residue accumulation on beams of order a millimeter in length. Care must be taken when devices are cleaned since the cleaning process may introduce further residue, or roughness (see example of silicon dipped in HF in Ref.~\cite{Klitsner-residue}).  

Table~\ref{tbl:Data} summarizes the silicon beam geometries and conductances shown in Fig.~\ref{fig:G-AoL}. The exponent $\beta\equiv n-1$ is inversely proportional to $G$, indicating a transition from diffuse to ballistic transport. Low $G$ devices have boundary-limited transport, and $\beta\simeq3$. High $G$ devices are approaching ballistic transport, and $\beta<3$ in the transition region. 

\begin{table}[htdp]
\caption{Conductance fit parameters $\alpha\equiv nK$ and $\beta\equiv n-1$ as a function of beam geometry, $G=\alpha\,T^{\beta}$. The entries marked with an asterisk indicate beams with smooth top surfaces (lift-off devices described in the text). }
\begin{center}
\begin{tabular}{c|c|c|c|c}
\hline
\hline
\,Length \,	&	\,Width \,  & 	\,$\alpha$ \,	& 	\,$\beta$\, 	&	\,G(150mK)  \,\\
($\mu$m) &	($\mu$m) & (pW/K$^{\beta+1}$) &				&		(pW/K) \\
\hline
5000	 & 35	 & 1054	 $\pm$ 55	 & \,3.13	 $\pm$ 0.06\,	 & 2.78 \\ 
5000	 & 45	 & 1449	 $\pm$ 50	 & 2.96	 $\pm$ 0.04	 & 5.28 \\ 
3000	 & 25	 & 499	 $\pm$ 36	 & 2.92	 $\pm$ 0.08	 & 1.96 \\ 
3000	 & 25	 & 1241	 $\pm$ 118	 & 3.27	 $\pm$ 0.09	 & 2.51 \\ 
3000	 & 25	 & 1210	 $\pm$ 103	 & 2.19	 $\pm$ 0.05	 & 18.99$^*$ \\ 
1500	 & 15	 & 625	 $\pm$ 44	 & 2.92	 $\pm$ 0.09	 & 2.46 \\ 
1500	 & 15	 & 583	 $\pm$ 31	 & 3.21	 $\pm$ 0.07	 & 1.32 \\ 
1500	 & 20	 & 2000	 $\pm$ 81	 & 2.38	 $\pm$ 0.03	 & 21.88$^*$ \\ 
1000	 & 25	 & 2406	 $\pm$ 37	 & 2.89	 $\pm$ 0.03	 & 10.00 \\ 
553	 & 20	 & 2521	 $\pm$ 67	 & 3.11	 $\pm$ 0.03	 & 6.90 \\ 
544	 & 20	 & 1922	 $\pm$ 132	 & 2.80	 $\pm$ 0.08	 & 9.48 \\ 
400	 & 24	 & 4292	 $\pm$ 386	 & 2.45	 $\pm$ 0.06	 & 41.10 \\ 
\hline
\hline
\end{tabular}
\end{center}
\label{tbl:Data}
\end{table}%

%%%%%%%%%%%%%%%%%%%%%%%%%%%%%%
%%%%%%%%%%%%%%%%%%%%%%%%%%%%%%
\section{Control of $G$ in Silicon-based detectors\label{sec:controlG}}

\begin{figure}[t]
\begin{center}
\includegraphics[width=8.5cm]{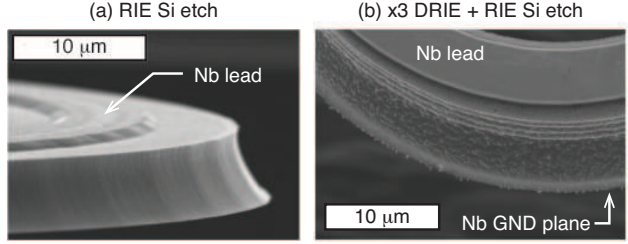}
\caption{SEM images of silicon beams. The Nb lead is 200 nm thick. (a) The isotropic RIE chemistry leaves smooth sidewalls with a slight taper. In Fig.~\ref{fig:G-AoL}, devices marked with crosses, squares, diamonds, and circles have similar sidewall profiles. (b) To minimize the top surface area roughened by the dry etch of the Nb lead layer, DRIE was used to define the beam width. The ridges at the top of the beam are scalloping from the DRIE cycles. Possible sources for the bright features on the sidewall are leftover passivation layer, and charged silicon features.  }
\label{fig:sidewall-SEM}
\end{center}
\end{figure}

In the design of prototype TES detectors, the choice of the beam length and width was guided by the data in Fig.~\ref{fig:G-AoL}. Several strategies were adopted in order to control the conductance and produce uniform detectors for the focal plane array described in Ref.~\cite{Eimer,Rostem}.  To minimize the top surface area of the silicon beams roughened by the dry etch, the width of the Nb leads were extended to cover a larger fraction of this surface, and the beam widths were defined with a combination of DRIE using the standard Bosch process~\cite{drie} followed by RIE to remove the silicon down to a Nb ground plane. The two processes are separated to reduce potential metallic contamination in the DRIE chamber while minimizing lateral silicon undercut from the isotropic RIE. The bottom surface of the beams in contact with the Nb ground plane are smooth, and scanning electron microscopy images of the sidewalls showed no significant features from the RIE etching of the beams employed in test structures,  Fig.~\ref{fig:sidewall-SEM}(a). The beam sidewalls were predicted to be nearly specular, with conductances comparable to the lift-off devices shown in Fig.~\ref{fig:G-AoL}. The top surface of the beams was expected to be slightly rough, $f\sim0.2$, since the area exposed during the dry etch step was reduced by a factor of two. Four beams 785 $\mu$m in length and 13 $\mu$m in width were integrated into the TES design.  

\begin{figure}[htbp]
\begin{center}
\includegraphics[width=8.5cm]{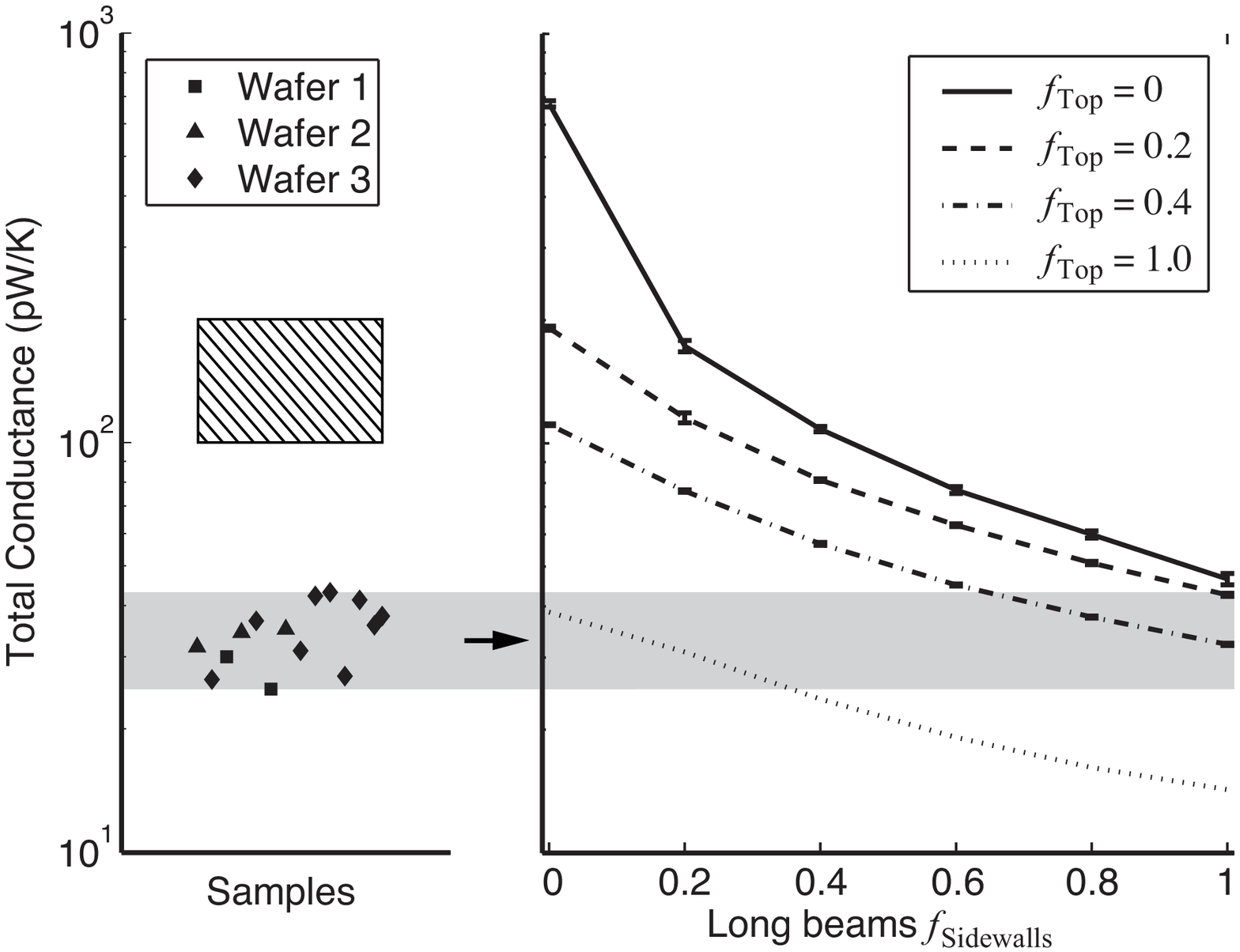}
\caption{(Left) Conductance of the four 785 $\mu$m long beams shown in Fig.~\ref{fig:Q-band-detector}. The measured conductance was a factor of $\sim$\,5 below the design conductance (hatched box). The cause of the variability is boundary-limited scattering. (Right) Simulations of the conductance as a function of beam wall diffuse fractions. The inferred diffuse fraction of the samples are moderately rough top surfaces, $f_{\rm Top}\le0.4$, and very rough sidewalls, $f_{\rm Sidewalls}\ge0.6$. }
\label{fig:LongBeam-Model}
\end{center}
\end{figure}

Post fabrication, the beams had well-defined top surfaces, but the sidewalls were roughened by the DRIE steps, Fig.~\ref{fig:sidewall-SEM}(b). The total conductance was a factor of 5 below the predicted target range. In principle, the scale length and height of the sidewall features that include the DRIE scalloping as well as passivation residue present a very diffuse surface for phonons above 100 mK. Figure~\ref{fig:LongBeam-Model} compares the total conductance of the four beams to simulations based on the diffusive-ballistic model described in Sec.~\ref{sec:model}. The diffuse fractions of the sidewall and top surface of the beams vary across a wafer and between wafers. 

The inferred diffuse fraction of the top surface of the beams is $f\leq 0.4$. A possible source for the moderate roughness is the 200 nm thick Nb lead sputter-deposited on each beam. The lead and beam are generally considered as decoupled systems, and the Nb conductance is estimated to be far below that of the silicon beam~\cite{Yefremenko1}. However, it is the diffusion process of the phonons in the silicon by the Nb layer that is of importance and generally unknown. It has been shown that even in the presence of large acoustic mismatch, phonons in silicon can be diffused at surfaces with metallic layers tens to thousands of Angstroms thick (see Fig.~8 in Ref.~\cite{Klitsner-residue} for examples of phonon diffusion caused by Al and Ti films below 1 K).  

\begin{figure}[htbp]
\begin{center}
\includegraphics[width=8.5cm]{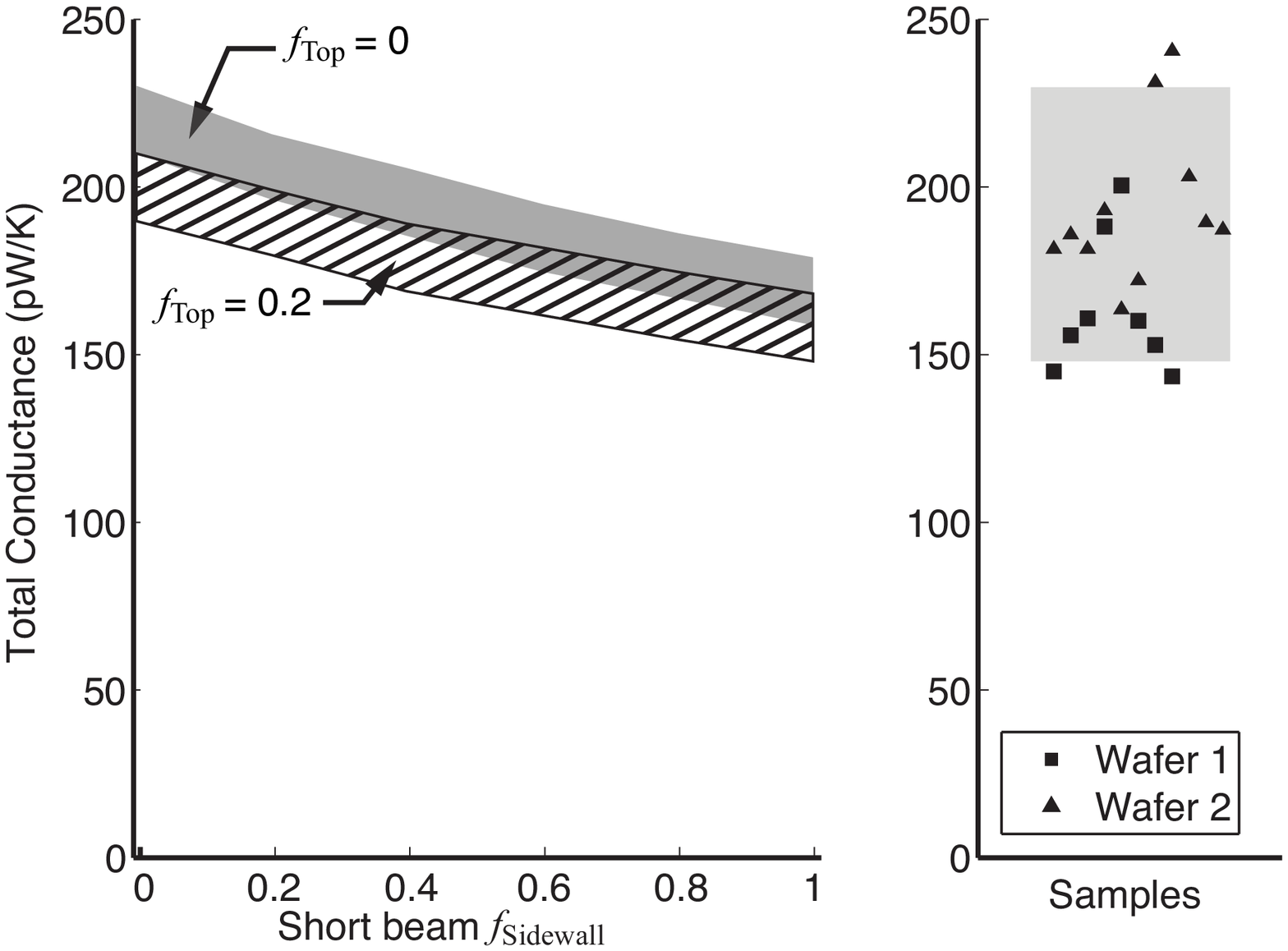}
\caption{(Left) Simulations of the total conductance of the beams including the short ballistic beam shown in Fig.~\ref{fig:Q-band-detector}. The shaded and hatched regions are defined by the measured range of conductance of the long diffuse beams shown in Fig.~\ref{fig:LongBeam-Model}. (Right) The measured conductances are within the predicted range of the simulations (light grey).   }
\label{fig:shortBeam}
\end{center}
\end{figure}

It is evident from the results in Fig.~\ref{fig:LongBeam-Model} that a solution to the problem of conductance control is a design that is least sensitive to the effect of fabrication conditions such as surface roughening and residue. Our approach was to reduce the length $L$ of a beam, which increases the phonon mean-free-path $\ell$ by reducing the interaction probability of the phonons with the rough sidewalls of the beam. The conductance is less sensitive to $\ell$ in the limit $L/\ell \le1$. Figure~\ref{fig:shortBeam} shows the simulated and measured total conductance when a ballistic beam 10 $\mu$m long and 13 $\mu$m wide is integrated into the detector as shown in Fig.~\ref{fig:Q-band-detector}. Given our knowledge of the surface roughness after the dry etch step (Fig.~\ref{fig:LongBeam-Model}), a diffuse fraction $f\le0.4$ was expected and subsequently obtained in practice. 

The total conductance, which is largely set by the ballistic beam, was predicted prior to fabrication. The diffuse beams merely support microstrip and bias leads necessary for the operation of the detector. However, they contribute the largest fraction to the variance in the total conductance across a wafer and between wafers. The magnitude and variance in $G$ of these beams were estimated from the data in Fig.~\ref{fig:LongBeam-Model}. The fractional standard deviation in total conductance is $\pm$8\% and within the requirement of the target instrument~\cite{Rostem}. 

The range of conductances that can be accessed with a ballistic beam is suitable for low-noise bolometric detectors operating at ground-based or air-borne telescopes where the signal power is of order 1-3 pW/K. In the ideal design, a single ballistic beam sets the conductance of the detector and carries all signal and bias leads. This approach is very well suited to the fabrication of kilo-pixel arrays, especially when the signal is coupled to the detector with free-space absorbers. For space applications where the signal power is less than 1 pW, the ballistic beam can be a phononic structure with phonon bandgaps that can be engineered to reduce the conductance~\cite{Maldovan, Zen}.

%%%%%%%%%%%%%%%%%%%%%%%%%%%%%%
%%%%%%%%%%%%%%%%%%%%%%%%%%%%%%
\section{Control of $C$ in Silicon-based detectors\label{sec:controlC}}

The heat capacity of the dielectric also plays an important role in the performance of thermal detectors. For a bolometer and calorimeter, $C$ affects the detector time constant, and in some cases anomalous heat capacity associated with the dielectric has been reported to increase thermodynamic noise of these detectors~\cite{excess-noise}. 

The heat capacity of the silicon membranes were determined from $C=G\,\tau_{th}$, where the thermal time constant $\tau_{th}$ was measured from the decay in temperature of a silicon membrane to a pulse in heater power. Compared to the Debye specific heat~\cite{Kittel}, $C_{\rm V}^{\rm Debye}/C_{\rm V}^{\rm Meas}=0.2$ at 200 mK and 0.7 at 1 K. The systematic difference between the measurements and the Debye approximation is due to the large contribution to the total heat capacity of the devices from the normal metal Au heater/thermometer and Pd-Au termination on the silicon membranes. Nonetheless, the measured specific heat are extremely small. For the detectors shown in Fig.~\ref{fig:Q-band-detector}, the estimated silicon heat capacity is $\sim$2 fJ/K at 150 mK, which is advantageous since the overall detector heat capacity can be controlled with a normal metal that can be reliably deposited on the membrane. We have successfully used Pd to achieve a heat capacity of 2 pJ/K~\cite{Rostem}. Measurements indicate that the specific heat of Pd below 1 K is 5.8 mJ mol$^{-1}$ K$^{-2}$, which is lower than the room temperature value of 9.4 mol$^{-1}$ K$^{-2}$~\cite{Kittel}. We have successfully deposited 400 nm Pd layers with a lift-off process. 

It is important to emphasize that there is no evidence for anomalous specific heat in the single-crystal silicon device layers. This is in contrast to the large excess specific heat of amorphous SiN$_x$ that is widely used in the fabrication of thermal detectors~\cite{Rostem-SiN,Zink,excess-noise}. 

%%%%%%%%%%%%%%%%%%%%%%%%%%%%%%
%%%%%%%%%%%%%%%%%%%%%%%%%%%%%%
\section{Conclusions}

The thermal conductance of single-crystal silicon devices can be precisely controlled using a short beam with ballistic-dominated phonon transport. The conductance of the short beam is largely set by its area and is insensitive to the details of the fabrication conditions that can vary over time. This approach is superior to the conventional use of long beams for the control of thermal conductance. Phonon transport in long beams is diffusive-ballistic and very sensitive to the detailed surface physics of the beams. The conductance is thus boundary-limited, and has been observed in beams with modest surface roughness (5-10 nm rms in surface height). Resonant phonon scattering has also been observed at 300 mK in beams with highly pitted surfaces with pit depths of order 30 nm. Boundary-limited scattering has contributed to a large variance, up to a factor of 5, in the conductance of devices fabricated and tested during this work. In contrast, when a beam with ballistic-dominated thermal transport is integrated into the design, the uniformity in conductance is reduced to a fractional deviation of $\pm$8\% in devices fabricated across two wafers, and this variability is largely set by the conductance of diffuse beams that support electrical and microwave leads to the device. In addition, we have found no evidence of excess specific heat in single-crystal silicon membranes. Hence, the total heat capacity of the detector can be effectively controlled with a normal metal film. For the transition-edge sensors (TESs) described in this work, the heat capacity was determined by a Pd layer 400 nm thick. Hence, the strategies outlined for the precision control of thermal conductance and heat capacity are well suited to the fabrication of uniform large-format arrays of TESs with sensitivities approaching 10$^{-18}$ W/Hz$^{1/2}$.

%%%%%%%%%%%%%%%%%%%%%%%%%%%%%%
%%%%%%%%%%%%%%%%%%%%%%%%%%%%%%
\section{Acknowledgement}

We gratefully acknowledge financial support from the NASA ROSES/APRA program. K. Rostem was supported by the NASA Postdoctoral Program at the Goddard Space Flight Center during the initial stages of this work. We thank Samelys Rodriguez for wirebonding and cryo-cable support. We thank our collaborators A. Ali, J. Appel, and T. A. Marriage for their contribution to the data acquisition. We thank the National Institute of Standards and Technology for providing the SQUID series array and time division multiplexer chips. 

\appendix*
\section{\label{sec:app:A} Radiative phonon transport}

We start from the usual definition of the infinitesimal view factor $dF_{dA_i-dA_j}$~\cite{Modest},
\begin{eqnarray}
dA_i\,dF_{dA_i-dA_j} = \frac{\cos\theta_i\cos\theta_j}{\pi\,|{\vec R}|^2}\,dA_j\,dA_i,
\label{eqn:VF}
\end{eqnarray}
where ${\vec R}$ is the vector between the infinitesimal surfaces $dA_{i}$ and $dA_{j}$, and $\theta_{i,j}$ are the angles between the normal vectors to $dA_{i,j}$ and ${\vec R}$ (see Fig.~\ref{fig:ModelDiagram}). We define an equivalent view factor for convenience, 
\begin{eqnarray}
F_{dA_i-dA_j} = dA_i\,dF_{dA_i-dA_j},
\end{eqnarray}
and note that the heat exchanged between the infinitesimal surfaces is
\begin{eqnarray}
dP_{dA_i-dA_j}(x) = q\,[T_i^4(x)-T_j^4(x)]\,F_{dA_i-dA_j},
\end{eqnarray}
where $T_{i,j}(x)$ are the temperatures of the surface elements. The total heat flux is the integral of $dP_{dA_i-dA_j}$ over the beam length,
\begin{eqnarray}
P = \int_{\mathcal{S}} dP_{dA_i-dA_j}(x),
\end{eqnarray}
where $\mathcal{S}$ represents the surfaces of the waveguide, which in our case is a rectangular beam. 

The equivalent view factor between any two perpendicular surface elements of the beam is
\begin{eqnarray}
\label{eqn:perp}
F_{dA_i-dA_j}^\perp  &=& \left\{ \frac{y_i\,z_j}{\pi[(x_i-x_j)^2+z_j^2+y_i^2]^2} \left[ \,f_m\prod_k (1-f_k)^{N_k}\right] \right\} \,dA_i\,dA_j, \\
		&=& \mathcal{F}^\perp (m; x_i,y_i,y_j,z_j) \,dx_i\,dy_i\,dy_j\,dz_j, \nonumber	
\end{eqnarray}
where the indices $m$ and $k$ sum over all surfaces of the beam. For parallel surface elements, 
\begin{eqnarray}
\label{eqn:para}
F_{dA_i-dA_j}^\parallel  &=& \left\{ \frac{t^2}{\pi[(x_i-x_j)^2+(y_i-y_j)^2+t^2]^2} \left[ \,f_m\prod_k (1-f_k)^{N_k}\right] \right\} \,dA_i\,dA_j, \\
		&=& \mathcal{F}^\parallel (f_m; x_i,y_i,y_j,z_j) \,dx_i\,dy_i\,dx_j\,dy_j. \nonumber
\end{eqnarray}
$N_k$ is the number of specular reflections experienced by phonons reflecting off the element $dA_j$ first, and then all other surfaces until a reference plane (e.g. $x=0$ in Fig.~\ref{fig:ModelDiagram}(a)) is reached. $N_k$ can be determined from the geometry of the specular scattering as described in Fig.~\ref{fig:ModelDiagram}(c), and is the sum of scattering events from each surface. For example, if the radiating element is on the top surface, $N_k$ can be found for phonons reflecting off the top and bottom surfaces, and reflecting off the sidewalls only. In this way, a different diffuse fraction can be assigned to each surface. It is also useful to note that the linear dependence of the equivalent view factors on $f_m$ can be translated to a linear dependence of the temperature gradient on the diffuse fraction when Eq.~\ref{eqn:perp} and~\ref{eqn:para} are finally substituted into Eq.~\ref{eqn:Px01}. 

The integral equations of the equivalent view factors over the surfaces of the beam are 
\begin{eqnarray}
F^\perp (f_m; x_i)\,dx_i  &=& dx_i\int \int \int \mathcal{F}^\perp (f_m; x_i,y_i,y_j,z_j) \,dy_i\,dy_j\,dz_j, \\
F^\parallel (f_m; x_i)\,dx_i &=& dx_i\int \int \int  \mathcal{F}^\parallel (f_m; x_i,y_i,y_j,z_j) \,dy_i\,dx_j\,dy_j,
\label{eqn:dfsdf}
\end{eqnarray}
where the limits of the integrals depend on the surface $m$ from which the phonons were originally radiated. 

Since a temperature gradient is only present in the principal direction of heat flow (x-axis), all surface elements at the plane of constant $x$ are at the same temperature and radiate the same flux. Hence, the equivalent view factor for the direct (non-scattered) heat flux $F^D(f_m; x_i)$ from the surface to the reference plane can be calculated analytically from Eq.~\ref{eqn:VF} by considering the view factor of a infinitesimal strip of width $w$ to the reference plane at $x=0$~\cite{Modest}.

\clearpage
%%Bibliography at end
%%\bibliography{report}

%%%%%%%%%%%%%%%

\end{document}